\documentclass[conference]{IEEEtran}
\IEEEoverridecommandlockouts

\usepackage{subcaption}  
\usepackage{color}
\usepackage{pifont}
\usepackage{bbm}
\usepackage{stfloats}
\usepackage[short]{optidef}
\usepackage{multirow}
\usepackage{comment}
\usepackage[table]{xcolor}

\usepackage{amsmath,amssymb,amsthm}
\usepackage{ragged2e} 

\usepackage{array}
\IEEEoverridecommandlockouts

\usepackage{cite}
\usepackage{amsmath,amssymb,amsfonts}
\usepackage{algorithmic}
\usepackage{graphicx}
\usepackage{textcomp}
\usepackage{xcolor}
\usepackage{amsmath}

\def\BibTeX{{\rm B\kern-.05em{\sc i\kern-.025em b}\kern-.08em
    T\kern-.1667em\lower.7ex\hbox{E}\kern-.125emX}}
\begin{document}
%
\title{Reliable Task Offloading in MEC through Transmission Diversity and Jamming-Aware Scheduling
}

\author{\IEEEauthorblockN{Ghazal Asemian, Mohammadreza Amini, Burak Kantarci}
\IEEEauthorblockA{\textit{School of Electrical Engineering and Computer Science} \\
\textit{University of Ottawa}\\
Ottawa, ON, Canada \\
Emails: \{gasem093, mamini6, burak.kantarci\}@uottawa.ca}
}

\maketitle

\begin{abstract}
Mobile Edge Computing (MEC) enables low-latency applications by bringing computation closer to the user, but dynamic task arrivals and communication threats like jamming complicate reliable task offloading and resource allocation.
In this paper, we formulate a dynamic MEC framework considering the transmission diversity that jointly addresses task scheduling and resource block (RB) assignment in the presence of jamming. First, we define and evaluate key network metrics—including dropped task ratio and bandwidth utilization—while maintaining service continuity by accounting for the existing commitments of the edge server to previously offloaded tasks. Then, we propose a jamming-aware offloading and RB allocation framework that leverages transmission diversity and optimal scheduling across distributed gNBs. The proposed solution is compared to a similar scenario without transmission diversity and two baseline strategies of first-come-first-served (FCFS) and shortest task first (STF). The proposed algorithm effectively mitigates the impact of jamming while enhancing resource utilization and minimizing task drop rates, making it highly suitable for mission-critical MEC applications. At signal-to-jamming-and-noise ratio (SJNR) of $4\ dB$, the proposed method achieves a $0.26$ task drop rate, outperforming the scenario without transmission diversity with a task drop rate of 0.50 and STF and FCFS strategies with 0.52 and 0.63 task drop rates, respectively.
\end{abstract}


\vspace{-3mm}
\section{Introduction} \label{Sec:Introduction}

The growing demand for latency-sensitive applications requires computing frameworks capable of real-time processing with strict quality of service (QoS) requirements \cite{3gpp.38.913}. Multi-access edge computing (MEC) brings computational resources closer to end-users, reducing latency and minimizing the load of centralized cloud infrastructure \cite{moshiri2025joint}. When integrated with 5G technologies such as multi-user Multiple-Input Multiple-Output (MU-MIMO), MEC further enhances network performance in terms of data rate, power efficiency, and delay reduction \cite{gao2021resource, ding2022hybrid, zhu2023multi}. Nonetheless, task offloading and resource scheduling remain the key challenges, as they must optimize performance while minimizing resource usage \cite{ghanem2022optimal, wang2023joint}.

Security in MEC-enabled networks represents another critical concern given the exposure of edge devices. Edge nodes are positioned closer to potential attackers \cite{ranaweera2021survey}. However, unlike cloud servers and data centers, edge devices do not benefit from strong security protocols due to their limited resources \cite{xiao2018security}.
Therefore, the presence of malicious entities, such as jamming, introduces another layer of complexity to task offloading and scheduling in MEC-enabled communication networks. Jamming attacks intentionally disrupt communication by injecting interfering signals by taking advantage of the shared and open nature of the wireless medium \cite{liu2024fight}. Under the jamming attack, the decoding error and task drop rates are increased, which degrades the overall reliability of the communication system \cite{roman2018mobile}. Moreover, the resulting rise in the number of dropped tasks necessitates more retransmissions, which leads to increased latency and resource consumption \cite{xu2021energy}.
To address these challenges, this work proposes a jamming-resilient framework for task offloading and resource block (RB) allocation in MEC systems. By leveraging transmission diversity (TD), considering edge server queue commitments, and optimizing scheduling decisions, the proposed approach mitigates the effect of jamming attacks while enhancing bandwidth efficiency and reducing task drop rate. The main contributions of this work are as follows:
\vspace{-1.5mm}
\begin{itemize}
    \item To enhance resource efficiency and scheduling effectiveness, we investigate critical network metrics, including task drop ratio and bandwidth utilization, while ensuring that task admission decisions respect the ongoing processing commitments of edge servers. This approach reduces the likelihood of dropping newly arrived tasks without compromising the execution of previously assigned ones.
    \item Propose a jamming mitigation framework for joint task offloading and RB allocation in a MEC environment. The framework leverages transmission diversity at the gNB and employs an optimal scheduling scheme to reduce the dropped task ratio while efficiently allocating RBs.

\end{itemize}


\vspace{-2mm}
\section{Related Works}
\label{sec: related}

\begin{table*}[htbp]
    \caption{\small Overview of task offloading techniques in literature and this work } \vspace{-2 mm}
    \label{table:offloading-compare}
    \centering
    \begin{tabular}{|c|c|c|c|c|c|}
    \hline
    \cellcolor{blue! 20} Ref. & \cellcolor{blue! 20} Technical Approach & \cellcolor{blue! 20} Attack Type & \cellcolor{blue! 20} KPI & \cellcolor{blue! 20} Server commitment & \cellcolor{blue! 20} Transmission Diversity \\
    \hline 
        \hline
    \cellcolor{gray! 20} \cite{xu2021energy} & Game-theory & Dynamic jamming & Energy consumption & \ding{55} & \ding{55} \\
    \hline
    \cellcolor{gray! 20} \cite{liu2021intelligent} & Hybrid (game-theory, DL) & Intelligent jamming & Defense utility & \ding{55} & \ding{55}\\
    \hline
    \cellcolor{gray! 20} \cite{li2024determinacy} & TLD & DoS & Overdue loss & \ding{55} & \ding{55}\\
    \hline
    \cellcolor{gray! 20} \cite{amini2025joint} & Genetic algorithm & On-off jamming & drop ratio, latency & \ding{55} & \ding{55}\\
    \hline
    \cellcolor{gray! 20} \cite{xiao2018security} & RL & Jamming & SINR, delay, energy consumption & \ding{55} & \ding{55}\\
    \hline
    \cellcolor{gray! 20} \cite{xiao2020reinforcement}& DRL & Smart jamming & Data rate, energy consumption, latency & \ding{55} & \ding{55}\\
    \hline
    \cellcolor{gray! 20} \cite{liu2024fight} & HDRL & Reactive jamming & Delay and energy & \ding{55} & \ding{55}\\
    \hline
    \cellcolor{gray! 20} This paper & Genetic Algorithm & Jamming & Drop and bandwidth utilization & \checkmark & \checkmark \\
    \hline
    \end{tabular}
    \vspace{-4mm}
\end{table*}

Significant research effort is dedicated to jamming mitigation frameworks in wireless communication systems. However, its impact on MEC systems and the development of corresponding countermeasures remain relatively underexplored.
An example of incorporating the game-theoretic approach for task offloading under a jamming attack can be found in \cite{xu2021energy}. Another example is discussed in \cite{liu2021intelligent}, integrating deep learning with game theory for power allocation against jamming.
Li et al. ~\cite{li2024determinacy} present a two-layer decomposition algorithm (TLD)to optimize task offloading under link-based Denial of Service (DoS) attack.
Amini et al. ~\cite{amini2025joint} formulates a multi-objective task offloading problem under on-off jamming and solves it using a genetic algorithm (GA).
With reinforcement learning (RL) gaining attention for anti-jamming methods in MEC, \cite{xiao2018security}, \cite{xiao2020reinforcement}, and \cite{liu2024fight} apply RL-based techniques for task offloading and scheduling, using Markov Decision Process (MDP), deep RL (DRL) with actor-critic and DQN structures, and hierarchical DRL (HDRL) with DDQN and TD3, respectively.
Unlike prior work, our proposed framework jointly minimizes task drop rate and bandwidth utilization under jamming attacks. We employ transmission diversity, server queue history, and optimal RB scheduling to address task failures caused by both communication degradation and poor scheduling, while accounting for the ongoing processing commitments of edge servers. The algorithm shows strong resilience against jamming, achieving a task drop rate as low as $0.37$ and bandwidth utilization of $0.5$ under severe conditions (signal-to-jamming-and-noise ratio (SJNR) of 0 dB).

\section{System Model}
\label{Sec:System_model}
    Consider a cellular network with multiple users connected to a gNodeB (gNB). The gNB is equipped with a computational server capable of efficiently processing user tasks. A MU-MIMO architecture with beamforming capabilities is considered, enabling the division of the serving area into multiple beam sectors. Each sector can simultaneously serve one or more users, improving spatial resource utilization and signal quality.

\vspace{-2mm}
\graphicspath{{pics/}}
\begin{figure}[htp]
    \centering
    \includegraphics[trim={8.15cm 0.5cm 7.5cm 4.45cm},clip , width=0.85\linewidth]{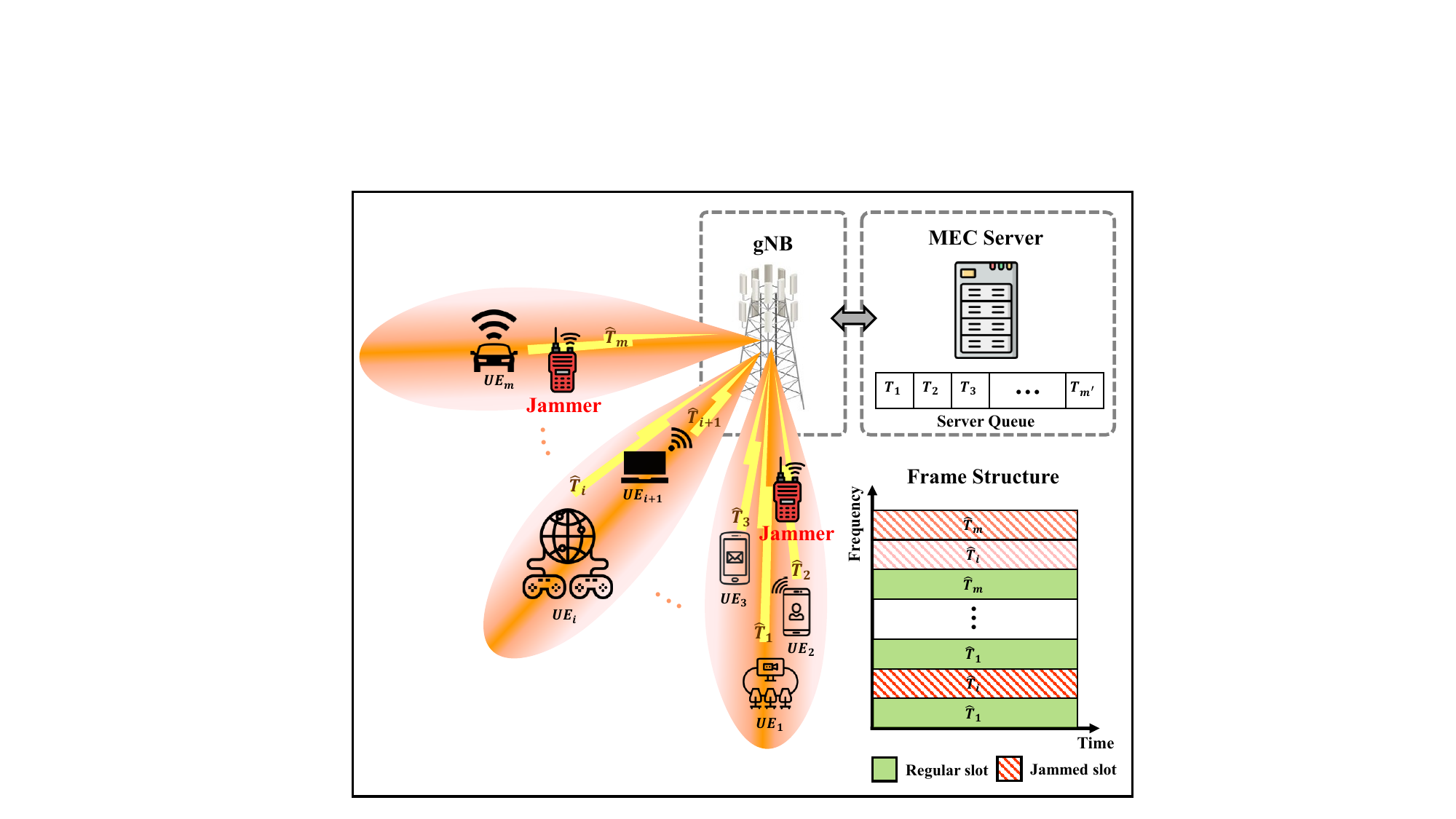}
    \caption{\small MU-MIMO MEC system under jamming attack.}
    \label{fig:scenario}
    \vspace{-5mm}
 \end{figure}

Assume that at time step $n$, there exists a set of tasks from various users that are to be offloaded to the gNB for processing at the server. Let $\hat{\mathcal{T}}^n=\{ \hat{T}^n_1, \hat{T}^n_2, \hat{T}^n_3,...,\hat{T}^n_m\}$ be the set of the mentioned tasks \footnote{For notational simplicity, we drop the time index $n$, as the rest of the formulation holds true for all $n$}. A typical task-say $i^{th}$ task-is characterized by a tuple $\hat{T}_i=\left( \hat{d_i}, \hat{b}_i, \hat{l}_i \right)$ in which $\hat{d}_i$ corresponds to the deadline, $\hat{b}_i$ represents the size of task in bits, and $\hat{l}_i$ indicates the computational load in CPU cycles. At this step, the server has a certain number of tasks in its queue (denoted by $\mathcal{T}=\{T_1, T_2, T_3,..., T_{m'}\}$), waiting to get processed. 
The communication environment is subject to interference from multiple jammers that attempt to disrupt the communication links by injecting jamming power. Consistent with 5G standards, communication between users and the gNB is carried out in the time-frequency domain, where time is split into slots and the total bandwidth is partitioned into several RBs. Each time slot contains $r_b$ RBs in the frequency domain. Due to beamforming capability and MU-MIMO architecture, along with the spatial distribution of users and jammers, the gNB experiences a specific SJNR for each transmitted task in $\hat{\mathcal{T}}$ at each RB.
Fig. \ref{fig:scenario} illustrates the considered MU-MIMO-based MEC system under multiple jamming attacks. The network includes multiple user equipment (UEs) connected to a gNB equipped with a local MEC server. Tasks generated by UEs are offloaded to gNB for processing, forming two queues. One for new arrival tasks ($\hat{\mathcal{T}}$) and another one for already existing tasks in the server queue ($\mathcal{T}$). Due to beamforming capabilities, the serving area is divided into beam sectors, where multiple UEs can be served. Meanwhile, one or more jammers attempt to disrupt the communications by injecting interference across different RBs in the time-frequency domain. The bottom right of the figure demonstrates the frame structure representing the time-frequency domain over which communication between users and gNB happens. To increase the success rate of task decoding, TD is employed where the same tasks can be transmitted over multiple RBs within the same time slot. Furthermore, the power level of the jamming signal is not uniform across all targeted RBs. Some RBs are heavily attacked while others are slightly affected.

Let $\gamma_{i,j}$ be the SJNR at the $j^{th}$ RB $\left( j \in \{1,2,3,...,r_b\} \right)$ for $i^{th}$ arrival task, i.e. $\hat{T}_i$. Then, the SJNR is given by: 
\begin{equation}
    \gamma_{i,j} = \frac{S_{i,j}}{J_{j} + \sigma^2_{j}},
\end{equation}

where \( S_{i,j} \), \( J_{j} \), and \( \sigma^2_{j} \) denote the received signal power for the \( i^{\text{th}} \) arrival task, the received jamming power, and the noise power, respectively, all measured on the \( j^{\text{th}} \) RB. Therefore, each task at each RB experiences a specific decoding error probability at the gNB side. Let $P^e_{i,j}$ be the decoding error probability of $\hat{T}_i$ on the $j^{th}$ RB. Note that $P^e_{i,j}$ is determined based on the configured modulation and coding scheme (MCS) by the gNB Medium Access Control (MAC) scheduler. Then, the error probability for the task $\hat{T}_i$ with $\hat{b_i}$ bits when transmitted on $j^{th}$ RB is written as, 

\begin{equation}
P^T_{i,j}=\sum_{b'=1}^{\hat{b_i}}\binom{\hat{b_i}}{b'}\big(P^e_{i,j}\big)^{b'}\big(1-P^e_{i,j}\big)^{\hat{b_i}-{b'}}.
\end{equation}

Task characteristics, user-specific, and network-level configurations affect key timing parameters that impact critical key performance indicators (KPIs) such as task drop rate, average task completion time, and decoding error probability. The proposed framework addresses jamming in delay-sensitive offloading by minimizing task drops caused by server deadline violation and decoding failures from interference.

\vspace{-2mm}
\section{Problem Formulation}
\label{Sec: Problem Formulation}
Dropped tasks fall into two categories: those with decoding errors due to low SJNR and deadline-expired tasks due to improper scheduling. TD effectively reduces decoding failures, while appropriate scheduling strategies help minimize deadline violations.
Let $\mathcal{R}$ be the set of all RBs available at $n^{th}$ time step which $||\mathcal{R}||=r_b$. Furthermore, assume $\hat{\mathcal{R}}_i$ is the set of all RBs assigned to $\hat{T}_i$ where $||\hat{\mathcal{R}_i}||=\hat{r}_{b_i}$, ($\hat{r}_{b_i}<r_b$). Therefore, decoding error probability for $\hat{T}_i$ considering TD can be obtained as:
\begin{equation}
    P^{TD}_i=\prod_{j\in \hat{\mathcal{R}}_i} P^T_{i,j},
\end{equation}
where $\bigcup_{i\in \hat{\mathcal{T}}} \hat{\mathcal{R}}_i \subseteq \mathcal{R}$. To formulate a joint RB allocation and task offloading, two indicators are defined. Let $I^{sch}_i$ and $I^{dec}_i$ be the binary indicators showing that the $i^{th}$ task is successfully scheduled and decoded, respectively. 
\begin{equation}
    I^{dec}_i=
    \begin{cases}
        1 & i\in \mathcal{T}\\
        1-P^{TD}_i & i\in \hat{\mathcal{T}}
    \end{cases}
\end{equation}
\begin{equation}\label{eq: I schedule}
    I^{sch}_i=
    \begin{cases}
        1 & t^c_i<d_i \, \, and \, \,  Q_i \neq 0\\
        0 & t^c_i<d_i \, \, and \, \,  Q_i = 0\\
        0 & t^c_i>d_i
    \end{cases}
\end{equation}
with $t^c_i$ as the completion time of the $i^{th}$ task. For all $\hat{T}_i$, the completion time consists of two parts, communication time and computation time. Furthermore, $Q_i$ is the position of $i^{th}$ task in the server queue that is obtained as: 

\begin{equation}\label{eq:Q}
    Q_i=\sum_{q=1}^{m+m'}q X_{i,q}\ .
\end{equation}
Then, the expected value of dropped tasks is given by:
\begin{equation}
    E[N_D]=(m+m')-\sum_{i\in (\hat{\mathcal{T}}\cup\mathcal{T})} I^{sch}_i I^{dec}_i\ .
\end{equation}

To calculate $t^c_i$, two binary indices have been introduced here. $X_{i,q}$ which is the scheduling index equals $1$ when $i^{th}$ task scheduled in the $q^{th}$ position of the queue server. $Y_{i,j}$ is the RB index shows if $j^{th}$ RB is assigned to $i^{th}$ task. Note that a typical task can only take one position in the queue server ($\sum_qX_{i,q} \leq 1$), and one position cannot be assigned to more than two tasks ($\sum_iX_{i,q}\leq 1$).
For RB assignment, the constraints are $\sum_{i\in \hat{\mathcal{T}}} Y_{i,j} \leq 1$, and $\sum_{j=1}^{r_b}Y_{i,j}\leq r_b$. 

The required processing time to complete the $i^{th}$ task using a CPU with frequency of $f$ is formulated as $t^p_i = \frac{\hat{l}_i}{f}$. 
Furthermore, communication latency can be expressed as $L_i^{com}=\frac{\hat{b}_i}{R_i},\ i\in \hat{\mathcal{T}}$. 
Where $R_i$ is the transmission rate for task $i$ obtained from Shannon's formula as $R_i=\max_j\{Y_{i,j} B \log_2(1+\gamma_{i,j})\}$,
with B as the RB bandwidth. Note that the maximum rate among those assigned RBs to $\hat{T}_i$ is selected. The total computing delay for the $i^{th}$ task, $L_i^{cmp}$, is calculated based on all the processing time of previously-scheduled tasks and the $i^{th}$ task itself. Therefore (while $Q_i$ already defined in (\ref{eq:Q})),
\vspace{-2mm}
\begin{equation}
    L_i^{cmp}=\sum_{i' \in \{\mathcal{T}\cup \hat{\mathcal{T}}\}}\sum_{q=1}^{Q_i}X_{i',q} \times t^p_{i'},
\end{equation} Finally, $t^c_i$ in (\ref{eq: I schedule}) can be expressed as:
\begin{equation}
    t^c_i=
    \begin{cases}
        L_i^{cmp}+L_i^{com} & \text{if}\ i \in \hat{\mathcal{T}}\\
        L_i^{cmp} & \text{if}\ i\in \mathcal{T}
    \end{cases}
\end{equation}

To avoid assigning excessive RBs to the task by the MAC scheduler, another metric, bandwidth utilization, $\eta_{BW}$, is defined which is the ratio of the total number of assigned RBs to the total number of RBs. Therefore,
\begin{equation}
    \eta_{BW}=\frac{\sum_{i\in \hat{\mathcal{T}}}\sum_{j=1}^{r_B}Y_{i,j}I^{dec}_iI_i^{sch}}{r_b}=\frac{\sum_{i\in \hat{\mathcal{T}}}\hat{r}_{b_i}I^{dec}_iI_i^{sch}}{r_b}
\end{equation}
Note that $\hat{r}_{b_i}=\sum_{j=1}^{r_b}Y_{i,j}$ in the above formulation.
Then, the optimization problem can be defined as:
\vspace{-3mm}

\begin{equation*}
\hspace{-50mm}\underline{\text{\textbf{Optimization Problem}}}
\end{equation*}\vspace{-0.2in}
\begin{mini!}|l|[2]                   
    {X_{i,q}, Y_{i,j}}{ \lambda\ \frac{E[N_D]}{m+m'} + (1-\lambda)\ \eta_{BW}\label{eq: main}}{}{}
	\addConstraint{\sum_{i} X_{i,q}\leq 1,\quad }{\sum_{q} X_{i,q}\leq 1 \, \, \label{eq:X constraint}} 
    \addConstraint{\sum_{i\in \hat{\mathcal{T}}}Y_{i,j} \leq 1 ,\quad \sum_j Y_{i,j}\leq r_b}{ \,\label{eq:Y constraint}}
    \addConstraint{I^{sch}_i = 1 \quad \forall i\in \mathcal{T} }{ \,\label{eq:I sch constraint}}
\end{mini!}
Constraint (\ref{eq:I sch constraint}) ensures that all of the already scheduled tasks in the queue must not drop due to the new scheduling order.

\section{Numerical Results}
\label{Sec:Numerical_result}
\subsection{Configuration Setting}
To evaluate the performance of the proposed method, simulations are implemented and performed in MATLAB. The key simulation parameters are summarized in Table \ref{tab: simulation params}. Task characteristics $\left( \hat{d_i}, \hat{b}_i, \hat{l}_i \right)$ are independently generated for each user following a uniform distribution within the ranges specified in Table \ref{tab: simulation params} adapted from \cite{ alameddine2019dynamic, tan2021energy, azizi2022deco}. The communication system employs 16 QAM modulation, and task arrivals are modeled as a Poisson process with an average inter-arrival time equal to the duration of one frame. Decoding error probability is computed based on the SJNR $\gamma$, the modulation scheme (16 QAM), and the diversity order of $1$ \cite{proakis2008digital}. For simplicity, the SJNR is assumed to be similar across all users and RBs. 
\vspace{-1mm}
\begin{table}[!ht]
	\centering
	\caption{\small MU-MIMO-based MEC Simulation Parameters} 
    \label{tab: simulation params}
	\begin{tabular}{|c|c||c|c|} \hline
		\cellcolor{gray! 40}\textbf{Parameter}	&	\cellcolor{gray! 40}\textbf{Value}  &	\cellcolor{gray! 40}\textbf{Parameter}  &  \cellcolor{gray! 40}\textbf{Value}  \\
		\hline \hline
		$\hat{d}_i$ & $[140,200]\ ms$ & $\hat{b}_i$ & $[1,10]\ KB$ \\ \hline
		$\hat{l}_i$ & $[2,50]$ Mega cycles & $f$ & $1\ GHz$ \\  \hline
		$m$ & $5$ & $m'$ & $3$ \\ \hline	
        $B$ & $100$\ KHz& modulation & 16 QAM \\ \hline	
        $r_b$ & $[1, 15]$ & $SJNR$ & $[0,30]$ \\ \hline	
	\end{tabular} 
\end{table}
\vspace{-2mm}

The optimization problem in (\ref{eq: main}) is solved using a GA with the objective function of $f(x)=\lambda\ \frac{E[N_D]}{m+m'} + (1-\lambda)\ \eta_{BW}$ implemented via MATLAB's Global Optimization Toolbox. 
The GA configuration setup is hypertuned with the population size of $2000$, constraint and function tolerance of $10^{-30}$, and maximum number of generations of $500$.
The \textit{crossoverscattered} and \textit{mutationgaussian} are used for the crossover function and mutation function in MATLAB.
To ensure statistical reliability and convergence stability, each GA run is executed 300 times, and the final performance metrics are obtained by averaging the results across the runs.
\vspace{-2mm}

\subsection{Performance Evaluation}
\vspace{-1mm}
At the beginning of each time slot, a new set of tasks and their corresponding properties are randomly generated. To investigate the influence of SJNR and the maximum number of available RBs, two simulation scenarios are considered. 
In the first scenario, the maximum number of RBs is fixed at $r_b = 10$, while the SJNR is varied across the range discussed in Table \ref{tab: simulation params}. This is to isolate the impact of communication quality under jamming on system performance, allowing for a clearer discussion on decoding reliability and task drop behavior without the effect of RB limitation.
In the second scenario, the SJNR is fixed at SJNR of $\gamma=5\ dB$, and the maximum available RBs are varied. This highlights how bandwidth availability influences the ability of the system to schedule and offload tasks under a severe jamming attack.
The results obtained from both scenarios are compared against three baseline strategies of Proposed - NTD, which is a similar scenario with proposed method without considering TD (i.e., $\sum_j Y_{i,j}\leq 1$), the first-come-first-serve (FCFS) scenario, which assigns tasks based on arrival order, and the shortest task first (STF) strategy, which prioritizes tasks with he smallest processing time. 

\graphicspath{{figs/}}
\begin{figure*}
    \centering
    \begin{subfigure}{0.24\textwidth}
        \includegraphics[trim={8.85cm 4.45cm 11.25cm 3.6cm},clip, width=1\textwidth, height=4cm]{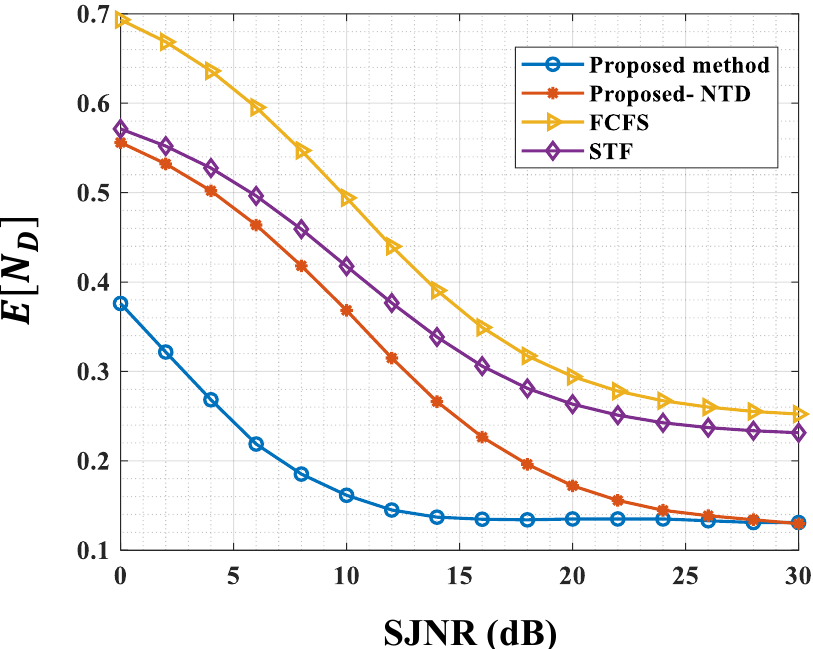}
        \caption{} 
        \label{fig:END vs sjnr}
    \end{subfigure}
    \hfill
    \begin{subfigure}{0.24\textwidth}
        \centering
        \includegraphics[trim={10.35cm 4.25cm 10cm 3.65cm},clip ,width=1\textwidth, height=4cm]{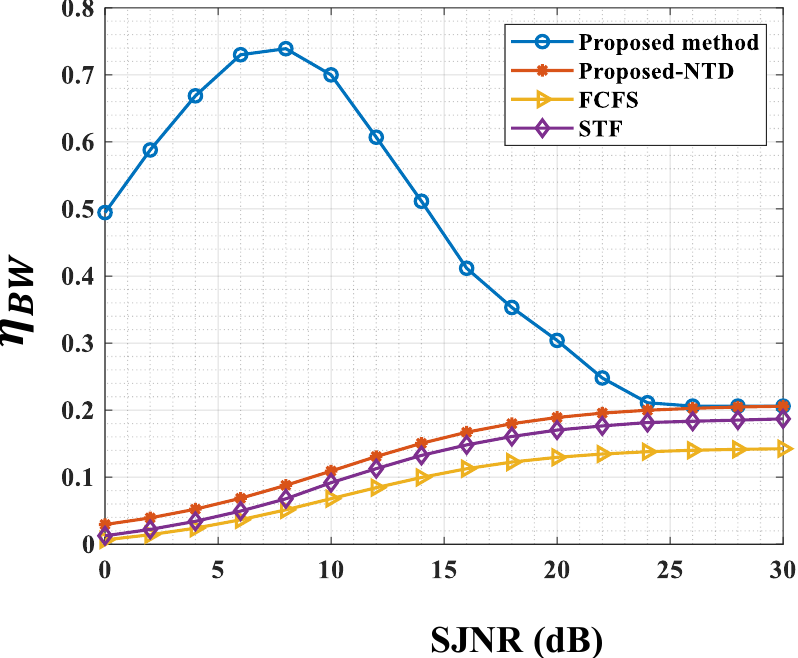}
        \caption{}
       \label{fig:eta vs sjnr}
    \end{subfigure}
        \begin{subfigure}{0.24\textwidth}
        \includegraphics[trim={8.9cm 4cm 11.25cm 4cm},clip ,width=1\textwidth, height=4cm]{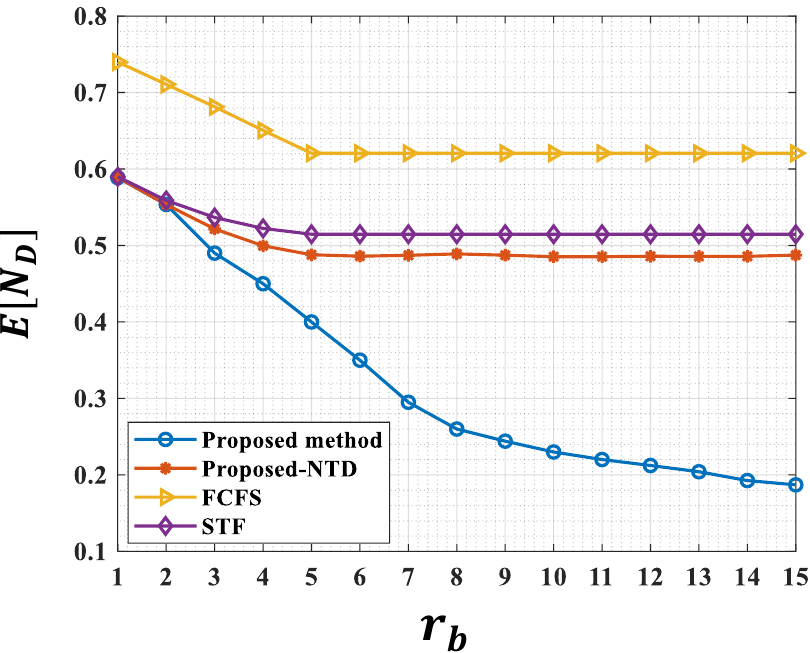}
        \caption{} 
        \label{fig:END vs rb}
    \end{subfigure}
    \hfill
   \begin{subfigure}{0.24\textwidth}
        \centering
        \includegraphics[trim={9cm 4.35cm 11.25cm 3.75cm},clip ,width=1\textwidth, height=4cm]{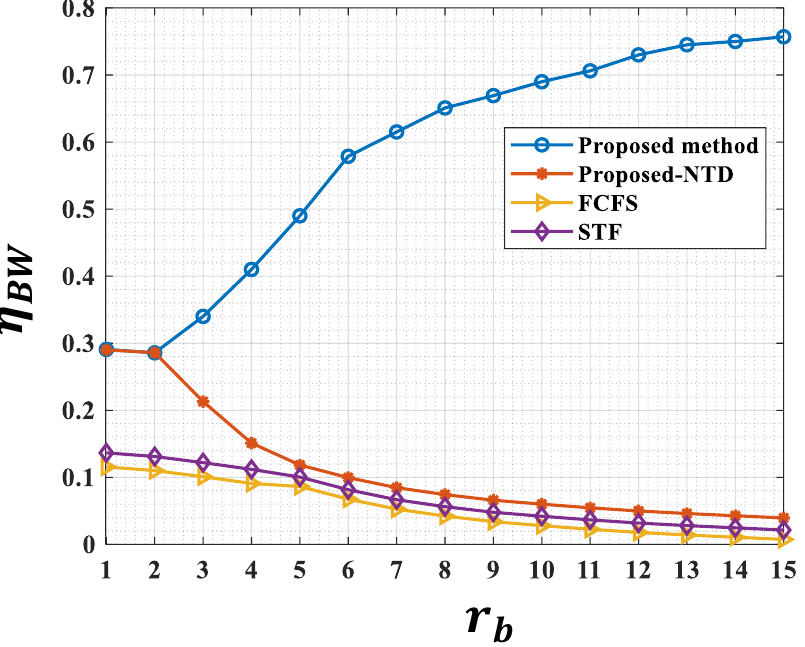}
        \caption{}
       \label{fig:eta vs rb}
       \vspace{-4mm}
    \end{subfigure}

   \caption{\small  (a) Drop task ratio versus SJNR for different task scheduling and RB allocation strategies- $r_b = 10$. (b) Bandwidth utilization versus SJNR for different task scheduling and RB allocation strategies- $r_b = 10$. (c) Drop task ratio vs. maximum available RBs for different task scheduling and RB allocation strategies- $\gamma =5\ dB$. (d) Bandwidth utilization vs. max available RBs for different task scheduling and RB allocation strategies- $\gamma = 5\ dB$. }
    \label{fig: performance eval}
    \vspace{-6mm}
\end{figure*}

Fig. \ref{fig:END vs sjnr} presents the drop rate based on the SJNR. The SJNR varies between $0\ dB$ to $30\ dB$ with a step size of 2. As shown in the figure, the task drop rate decreases with increasing SJNR for all strategies as the decoding probability increases and the number of dropped tasks due to decoding failure in response to jamming reduces. The proposed method represented by the blue curve performs significantly better than the other three methods, particularly in lower SJNR (harsher jamming conditions). The FCFS and STF, demonstrated with yellow and purple curves respectively, perform worse, as they lack any jamming mitigation strategy or transmission diversity. The proposed method, without considering TD, shown with the red curve, demonstrates the next best performance as it includes a higher decoding error due to the lack of transmission diversity. However, as the SJNR approaches $30\ dB$, both the proposed method and its variant without transmission diversity converge, due to the lower error probability and fewer retransmission requirements.

Fig. \ref{fig:eta vs sjnr} illustrates the impact of SJNR on bandwidth utilization. Under severe jamming conditions (i.e., low SJNR), the proposed method has the highest bandwidth utilization. This high utilization value arises from the increased number of RBs allocations required to deal with high decoding error probabilities, resulting in more retransmissions. As the jamming power decreases (i.e., SJNR improves particularly around $\gamma=8\ dB$), less tasks are expected to be erroneously decoded, which reduces the need for multiple retransmissions. Hence, fewer RBs are required for each transmitted task, reducing the overall bandwidth utilization. For other methods, the bandwidth utilization is low at low SJNR values due to a high task drop rate and limited resource usage. As the SJNR increases, $\eta_{BW}$ improves accordingly, eventually converging at high SJNR values where the rate of erroneous task decoding is decreased.

In figures Fig. \ref{fig:END vs rb} and Fig. \ref{fig:eta vs rb}, the impact of the maximum number of available RBs ($r_b$) on task drop rate and bandwidth utilization is investigated, respectively. As demonstrated in Fig. \ref{fig:END vs rb}, the proposed algorithm outperforms other strategies while the $r_b$ increases. When $r_b=1$, the proposed method resembles the scenario without TD, since only one RB is available for each task and there is no opportunity for retransmissions. As $r_b$ increases, the proposed algorithm is provided with additional flexibility to assign more RBs per task, enabling retransmission for the tasks affected by jamming. This effectively reduces the overall task drop rate. In contrast, the other three strategies have a significantly high drop ratio when $r_b$ is less than the number of arrival tasks, due to inefficient bandwidth to serve all tasks. As $r_b$ approaches $r_b=5$, which matches the number of incoming tasks per slot, the drop rate decreases for all methods. However, beyond this point, further increasing $r_b$ does not have any improvement for the drop rate. This is due to the fact that these strategies do not benefit from transmission diversity, thus, no retransmission is scheduled for dropped tasks. Nevertheless, the proposed algorithm without TD shows a better performance as the GA schedules the tasks in an optimal allocation strategy. A similar trend is observed in Fig. \ref{fig:eta vs rb}, in which at $r_b=1$ the proposed method performs similarly to Proposed-NTD as there are no available RBs for retransmissions. As $r_b$ increases, the algorithm leverages the additional available RBs for retransmission of tasks, which increases the bandwidth utilization. This trend continues until all tasks are successfully retransmitted and scheduled. 
For the other three strategies, when $r_b$ is less than the number of arrival tasks, with $r_b$ increasing, the number of scheduled tasks increases as well. For $r_b>5$, the total number of scheduled arrival tasks remains at the maximum number of arrival tasks while $r_b$ increases, which results in a gradual decrease in overall bandwidth utilization.
\vspace{-2mm}
\section{Conclusion}
\label{Sec:Conclusions}
This paper addresses dynamic task offloading and resource allocation in MEC environments under jamming attacks by integrating transmission diversity, server queue awareness, and optimized RB scheduling. 
The proposed framework reduces decoding errors and deadline violations, achieving lower task drop rates and higher bandwidth efficiency compared to FCFS, STF, and non-diversity baselines. Notably, under severe jamming conditions (for instance, SJNR of 4 dB), the proposed method achieves a task drop rate of 0.26. Our ongoing work explores advanced jamming prediction and DRL-based solvers to further improve the system's resilience in dynamic environments.

\vspace{-4mm}
\section*{Acknowledgment}
\vspace{-2mm}
This work was supported in part by funding from the Natural Science and Engineering Research Council (NSERC) CREATE TRAVERSAL and Discovery Programs.

\vspace{-2mm}
\bibliographystyle{IEEEtran}

\end{document}